\documentclass[11pt]{article}
\usepackage{moriond,epsfig}

\def\be{\begin{equation}}
\def\ee{\end{equation}}
\def\bea{\begin{eqnarray}}
\def\eea{\end{eqnarray}}

\begin{document}
\vspace*{4cm}
\title{DIFFRACTION : RECENT RESULTS AND IMPLICATIONS FOR LHC}

\author{ L. SCHOEFFEL }

\address{CEA Saclay, DAPNIA-SPP,\\
91191 Gif-sur-Yvette Cedex, France}

\maketitle\abstracts{
With the knowledge of diffractive parton densities extracted from HERA data,
we discuss the observation of exclusive events using the dijet
mass fraction as measured by the CDF collaboration at the Tevatron. In particular the
impact of the gluon density uncertainty is analysed. Some prospects are given for diffractive physics at the LHC.
}

\section{Diffraction at HERA}

\begin{figure}
\begin{center}
\includegraphics[totalheight=10cm]{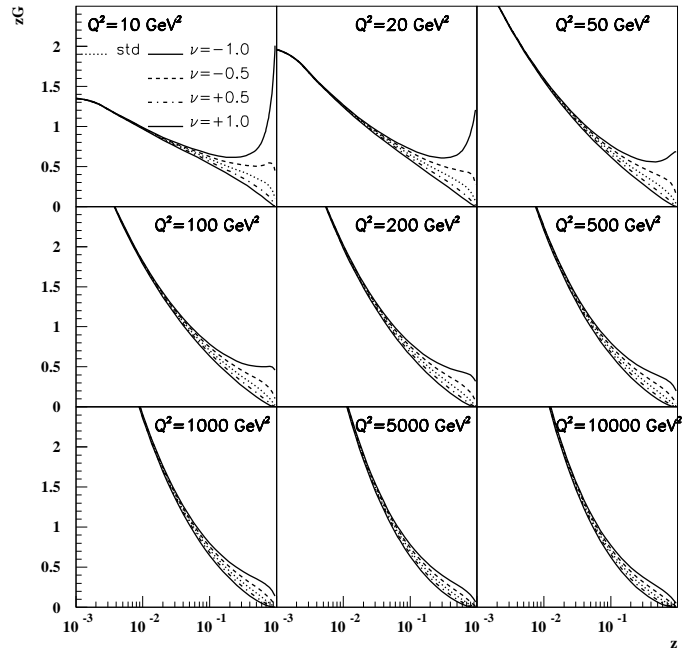}
\caption{Uncertainty of the gluon density at high $\beta$ (here $\beta\equiv z$).
The gluon density is multiplied by the factor $(1-\beta)^{\nu}$ where $\nu$=-1.,
-0.5, 0.5, 1. The default value is $\nu =0$.}
\label{gluon}
\end{center}
\end{figure}

Since years, the Pomeron remains a subject of many  interrogations. Indeed, 
defined as the virtual colourless carrier of strong interactions,  the nature of 
the Pomeron is still a real challenge. While in the perturbative 
regime  of QCD it can be defined  as a compound system of two  gluons 
in the approximation of resumming the leading logs in  energy, its 
non-perturbative structure is basically unknown.
In the recent years, an interesting experimental  investigation on 
``hard'' diffractive processes led to a new insight into  those problems. At 
the HERA accelerator, it has been discovered that  a non negligeable 
amount of $\gamma^*\!-\!{\rm proton}$ deep inelastic events can be produced with 
no 
visible breaking of the incident proton. There are various phenomenological 
interpretations of this phenomenon, but a very appealing one relies upon a partonic interpretation 
of  the structure of the Pomeron \cite{ingelman} . In fact, it is possible to nicely describe the 
diffractive cross-section data from HERA
 by  perturbative QCD evolution equations of  
parton distributions in the Pomeron combined with  flux factors 
describing phenomenologically the probability of finding a Pomeron state in the 
proton \cite{pdfs}.  Sets of  quark and gluon distributions in the Pomeron following 
these equations are obtained. 
The gluons dominate the
diffractive exchange and carry approximately 70 \% of the momentum. The diffractive gluon
density is presented in figure ~\ref{gluon}.
At high $\beta$, where $\beta$ denotes the 
fraction of the particular parton in the pomeron, 
this density is not well constrained from the QCD fits. 
To quantify this uncertainty, we multiply the gluon distribution by the
factor $(1 - \beta)^{\nu}$ as shown in figure ~\ref{gluon} : we obtain the uncertainty on the parameter $\nu$,
$\delta(\nu)=0.5$, which corresponds to a large spread at large $\beta$ for the gluon density.

In the following, 
we investigate how this uncertainty influences the
results on dijet mass fraction as measured at the Tevatron.

\section{Diffraction at Tevatron and LHC}

\begin{figure}
\begin{center}
\epsfig{figure=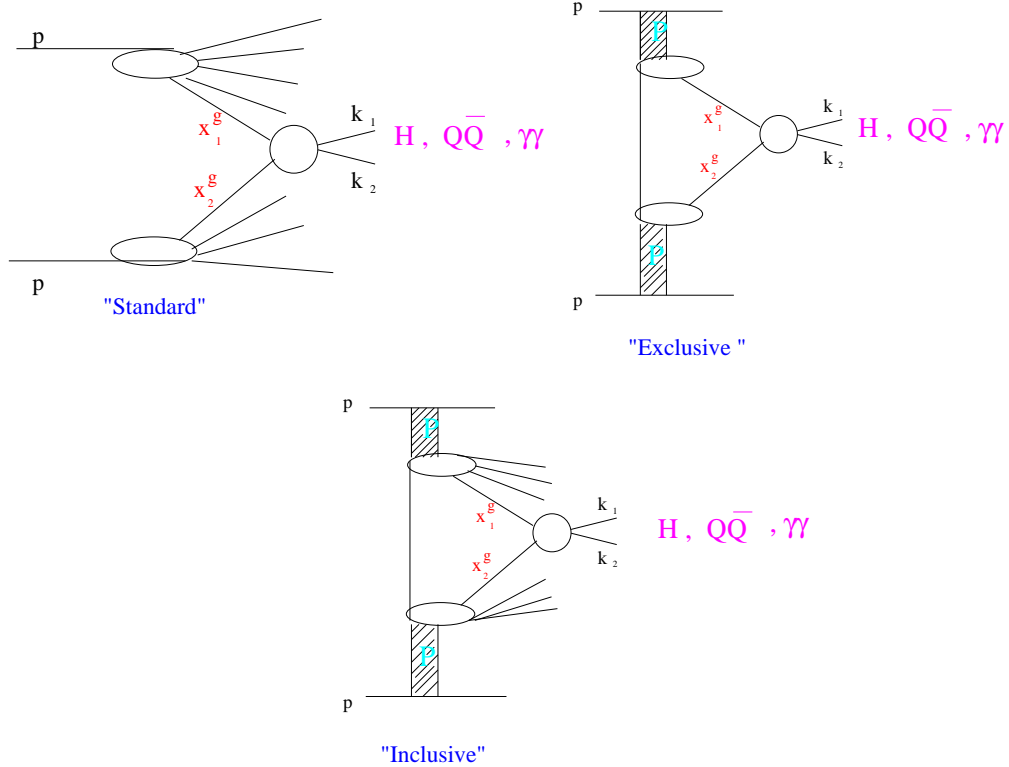,height=4.in}
\end{center}
\caption{\it Scheme of non diffractive, inclusive double pomeron exchange and 
 exclusive events at the Tevatron or LHC}
\label{ini}
\end{figure}

A schematic view of non diffractive, inclusive double pomeron exchange and 
exclusive diffractive events at the Tevatron or the LHC is displayed
in figure~\ref{ini}. The upper left plot shows the "standard" non diffractive events
where the Higgs boson, the dijet or diphotons are produced directly by a 
coupling to the proton associated with proton remnants. The bottom plot displays
the standard diffractive double Pomeron exchange (DPE) where the protons remain
intact after interaction and the total available energy is used to produce the
heavy object and the pomeron remnants.
These events can be described using the diffractive gluon density 
measured at HERA  and shown in figure \ref{gluon}. 
There may be a third class of processes displayed in
the upper right figure, namely the exclusive diffractive production. 
Exclusive events allow a precise reconstruction of the
mass and kinematical properties  of the central object
using the central detector or even more precisely  using very forward detectors
installed far downstream from the interaction point. 
The mass of the produced object can be computed using
roman pot detectors and tagged protons,
$
M = \sqrt{s \xi_1 \xi_2}
$,
where $\sqrt{s}$ is the energy of the reaction in the center of mass frame
and $\xi_{1,2}$ represent the fractions of energy lost by both protons. 
We see immediately the advantage of those processes : we can benefit from
the good roman pot resolution on $\xi_{1,2}$ to get a good resolution on mass. Therefore, it is
 possible to measure the mass and the kinematical properties of the 
produced object and use this information to increase the signal over background
ratio by reducing the mass window of measurement \cite{royon}. 

If such exclusive processes exist in DPE, 
the most appealing is certainly the Higgs boson
production through this channel at the LHC \cite{royon}. 
It cannot be observed at the Tevatron
due to the low production cross section, but one 
can use present measurements at the Tevatron to investigate any evidence for the existence of exclusive 
production in DPE.

\section{Dijet mass fraction at the Tevatron}

\begin{figure}[h]
\begin{center}
\includegraphics[totalheight=7cm]{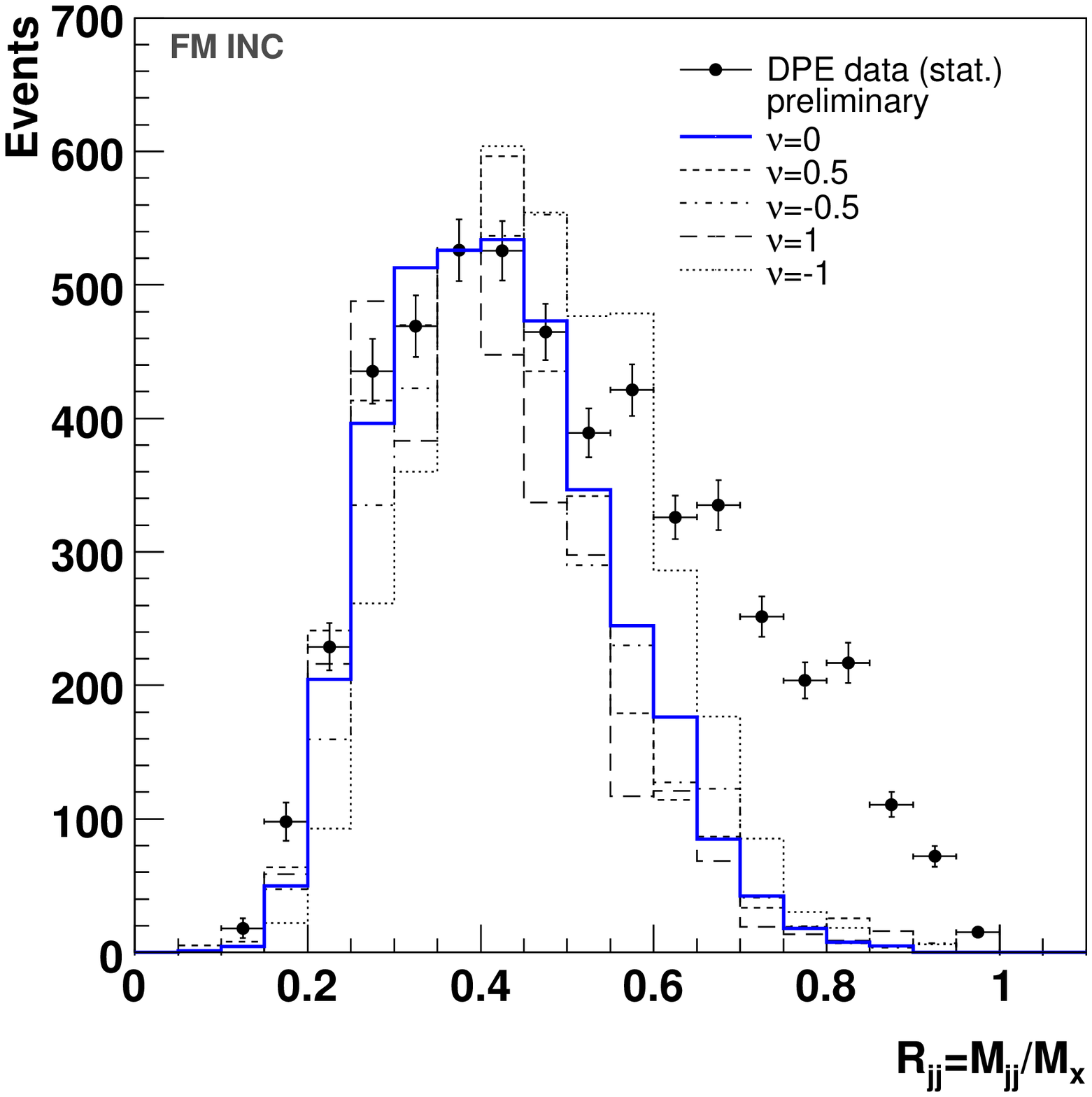}
\includegraphics[totalheight=7cm]{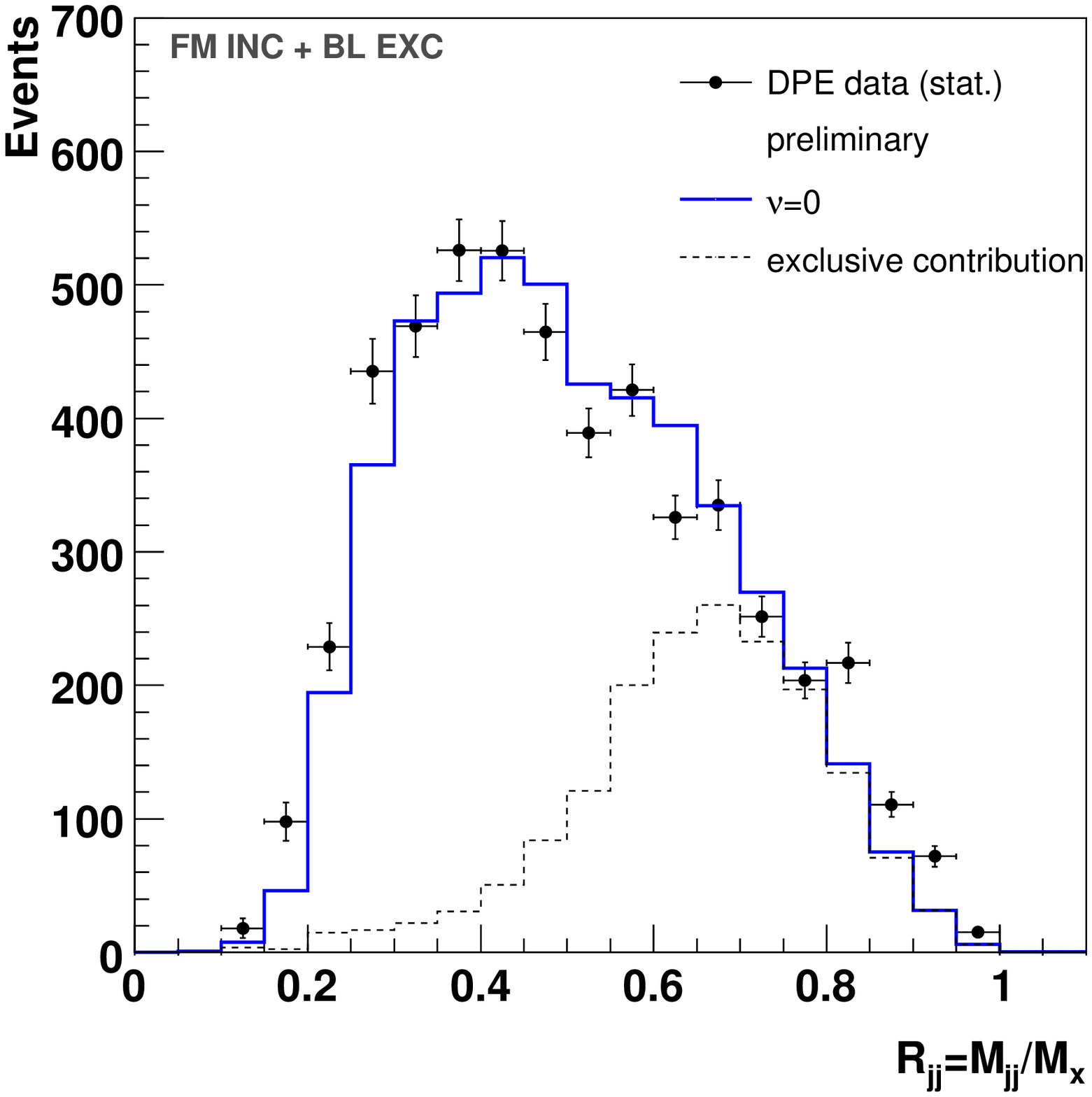}
\caption{Dijet mass fraction for jets $p_T>10\,\mathrm{GeV}$. 
The data are compared to inclusive model predictions including the
uncertainty of the gluon density at high $\beta$ (left) and to the sum of inclusive and
exclusive predictions (right).}
\label{dijetmass}
\end{center}
\end{figure}

The CDF collaboration has measured the so-called dijet mass fraction (DMF) in dijet
events when the antiproton is tagged in the roman pot detectors
 and when there is a rapidity gap on the proton side to ensure that the
event corresponds to a double pomeron exchange \cite{royon,proceedings}. 
The measured observable  $R_{JJ}$ is defined as the ratio of the mass carried by the two jets divided by the total
diffractive mass. 
The DMF turns out to be a very appropriate observable for identifying the exclusive 
production, which would manifest itself as an excess of the 
events towards $R_{JJ}\sim 1$. 
Indeed, for exclusive events, the dijet mass is essentially equal
to the mass of the central system because no pomeron remnant is present.
Then, for exclusive events, the
DMF is 1 at generator level and can be smeared
out towards lower values taking into account the detector resolutions.
The advantage of DMF is that one can focus on the shape of the distribution. The observation 
of exclusive events does not rely on the overall normalization which might be strongly dependent on
the detector simulation and acceptance of the roman pot detector.
Results are shown
in figure~\ref{dijetmass} with Monte-Carlo expectations calculated using DPEMC \cite{dpemc}. 
As we have seen in section 1, the uncertainty on the gluon density is large at large $\beta$,
which directly reflects in  different shapes for the DMF in the inclusive part \cite{olda}.
This is illustrated in figure~\ref{dijetmass} (left), where we show the impact of the parameter
$\nu$, which quantifies the diffractive gluon density error, on the shape of the DMF. We observe that
it is not sufficient to reproduce the behaviour of the DMF when $R_{JJ}\sim 1$.
Indeed,
we see a clear deficit of events towards high values of the DMF, 
where exclusive events are supposed to occur. 
In figure~\ref{dijetmass} (right), a specific model describing exclusive events \cite{bl} is added to
the inclusive prediction and we obtain a good agreement between data and the sum of MC expectations \cite{dpemc}.
It is a first evidence that exclusive events could contribute at the Tevatron.

\section{Dijet mass fraction at the LHC}

The search for exclusive events at the LHC can be performed in the same
channels as the ones used at the Tevatron. A direct precise measurement of
the gluon density in the pomeron through the measurement of the diffractive
 dijet cross section at the LHC will be necessary to study in detail the
exclusive events in the dijet channel and measure their cross section. This
is why it is important to have the roman pots and the Silicon detectors (inside roman pots)
installed during the 2009-2010 shutdown so that these measurements will
allow to tune the models and the MC. On the other hand,
it is also important to look for different methods to show the existence of
exclusive events \cite{royon}.
In addition, some other possibilities benefitting from the high luminosity
 of the LHC appear. One of the cleanest way to show the existence of
exclusive events would be to measure the dilepton and diphoton cross section
 ratios as a function of the dilepton/diphoton mass. If exclusive events
exist, this distribution should show a bump towards high values of the dilepton/diphoton
 mass since it is possible to produce exclusively diphotons but
not dileptons at leading order.

\section{Conclusions}

We have discussed a first evidence for the existence of exlusive events in double pomeron exchange at the Tevatron.
If such events can be also observed at the LHC, it would be possible to produce a Higgs boson as well as of a dijet system
regarding the cross section values accessible at the LHC.
The great benefit of exclusive events concerns the precise reconstruction of the
mass of the central object,
using roman pot detectors
installed far downstream from the interaction point \cite{royon}. It gives the opportunity to
work with a favorable  signal/background ratio compared to standard Higgs searches 
with a mass below  150 GeV.

\end{document}